\documentclass{article}
\usepackage{epsfig}
\usepackage{hyperref}
\input{epsf}
\def\r{{\tilde r}}
\def\b{{\tilde \phi}}
\title{Gravity of a static massless
scalar field and a limiting Schwarzschild-like geometry}
\author{M Gaudin$^1$, V Gorini$^{2,3}$, A Kamenshchik$^{2,4,5}$,\\
U Moschella$^{2,3,6}$ and V Pasquier$^1$}
\date{}
\begin{document}
\maketitle \hspace{-6mm}$^{1}$Service de Physique Th\'eorique, C.E.
Saclay, 91191 Gif-sur-Yvette, France
\\$^{2}$Dipartimento di Scienze
Fisiche e Mathematiche, Universit\`a
dell'Insubria, Via Valleggio 11, 22100 Como, Italy
\\$^{3}$INFN, sez. di Milano, Via Celoria 16, 20133 Milano, Italy
\\$^{4}$Dipartimento di Fisica and INFN, via Irnerio 46, 40126 Bologna, Italy
\\$^{5}$L.D. Landau Institute for Theoretical Physics,
Russian Academy of Sciences, Kosygin str. 2, 119334 Moscow, Russia
\\$^{6}$DASS, Universit\`a dell'Insubria, Via Dunant
2, 21100 Varese, Italy \vskip50pt

\begin{abstract}
We study a set of static solutions of the Einstein equations in
presence of  a massless scalar field and establish their
connection to the Kantowski-Sachs cosmological solutions based on
some kind of duality transformations. The physical properties of
the limiting case of an empty hyperbolic spacetime
(pseudo-Schwarzschild geometry) are analyzed in some detail.
\end{abstract}

\vskip50pt

The construction of spherically symmetric static solutions of
Einstein equations has been attracting the attention of researchers
since the time of the classical works by Schwarzschild
\cite{Schwarz}, Tolman \cite{Tolman} and Oppenheimer and Volkoff
\cite{Op}. In this note we study solutions describing a spacetime
filled with a massless scalar field and  their corresponding
limiting vacuum forms.

Consider first a static spherical metric of the following form
\begin{equation}
ds^2 = b^2(r) dt^2 - a^2(r) (dr^2 + d\theta^2 + \sin^2\theta
d\varphi^2). \label{metric}
\end{equation}
Einstein's equations are then written as follows
\begin{eqnarray}
&& G^t_t =  \frac{{a'(r)}^2 +{a(r)}^2  -
2\,a(r)\,a''(r)}{{a(r)}^4}= \varepsilon \label{EinK0s},\\
&& G^r_r = \frac{{a(r)}^2\,b(r) -b(r)\,{a'(r)}^2 -
2\,a(r)\,a'(r)\,b'(r)}{{a(r)}^4\,b(r)}= -\varepsilon \label{EinKrs} ,\\
&& G^\theta_\theta = G^\phi_\phi =  \frac{ b(r)\,{a'(r)}^2  -
a(r)\,b(r)\,a''(r) - {a(r)}^2\,b''(r)}{{a(r)}^4\,b(r)}
\label{EinKthets}= \varepsilon
\end{eqnarray}
where
\begin{equation}
\varepsilon = \frac{4\pi{\phi'}^2}{a^2}.\label{enKs}
\end{equation}
and $\phi = \phi(r)$ is a massless scalar field. It is convenient to
introduce the notations:
\begin{equation}
A \equiv \frac{{a}'}{a},\label{A-def} \;\;\;\;\;B \equiv
\frac{{b}'}{b}.\label{B-def}
\end{equation}
Einstein's equations then imply that
\begin{eqnarray}
 &&{A'} + A^2 +
AB - 1 = 0,\label{Comp}\\
&&{A'} + A^2 - {B'} - B^2 - 1 = 0.\label{Comp1}
\end{eqnarray}
By eliminating $A'$ one gets the following relationship:
\begin{equation}
A = -\frac{{B'}}{B} - B,\label{ABK}
\end{equation}
From Eq. (\ref{Comp}) we then see that $B^{-1}$ satisfies the
equation for  an upside-down harmonic oscillator:
\begin{equation}
\left(\frac1B\right)^{\prime\prime} - \frac1B = 0.\label{eqBs}
\end{equation}
The general solution for Eq. (\ref{eqBs}) can be written as follows
$$B =\frac 2{ce^{\,r} + de^{-r}},$$
which implies
$$A = 1 - \frac{2\,\left( d + e^r \right) }{d + c\,e^{2\,r}}$$
One can easily solve Eqs. (\ref{B-def}) for $a$ and $b$ and compute
the square of $\phi'$, which is pointwise proportional to the energy
density $\varepsilon$:
\begin{equation}
{\phi'}^2 = \frac{\varepsilon a^2}{4\pi} = -\frac{\left( 1 + c\,d
\right) }{\pi{\left(c\,e^{\,r} + d\,e^{-r}  \right) }^2}.
\label{edens0}
\end{equation}
If we demand the energy density to be nonnegative, we obtain that
the parameters $c$ and $d$ must satisfy the inequality $cd\leq-1$.
By introducing
\begin{equation}
\gamma^2 = {-\frac{1}{cd}},\;\;\;\;r_0 = -\log\sqrt{-\frac c d},
\end{equation}
we may parametrize the physically acceptable solutions as follows:
\begin{equation}
B = \frac{\gamma}{\sinh (r-r_0)}, \;\;\;\;\; A = \coth (r-r_0) -
\frac{\gamma}{\sinh (r-r_0)}\label{asols}
\end{equation}
where the parameter $\gamma$ must satisfy the restriction
\begin{equation}
\gamma^2 \leq 1\label{restrict}
\end{equation}
in order to insure positivity of the energy density. On the other
hand the parameter $r_0$ has no physical meaning and can be set
equal to zero. Explicit expressions for $a$, $b$ and ${\phi'}^2$ are
easily written:
\begin{equation}
ds^2 = \left(\tanh\frac{r}{2}\right)^{2\gamma} dt^2 -
\frac{a^2_0\sinh^2 r}{\left(\tanh\frac{r}{2}\right)^{2\gamma}} (dr^2
+ d\theta^2 + \sin^2\theta d\varphi^2),\label{bsols}
\end{equation} \begin{equation} {\phi'}^2 = \frac{\varepsilon
a^2}{4\pi} = \frac{\left( 1 - \gamma^2\right) }{4\pi \sinh^2 r}.
\label{edens}
\end{equation}
The values $\gamma = \pm 1$ correspond to the limiting cases of
empty spacetimes; the value $\gamma = 0$ corresponds to the
maximum possible value of the spatial variation of the scalar
field; in this case $b(r) = 1$ and the time coordinate $t$ is
everywhere ticking at the same rate.

Solutions of this kind have already been obtained in \cite{Xan} in
terms of isotropic coordinates. The metric (\ref{bsols}) can be
written in isotropic coordinates by the change of coordinates $ r =
\log \r $:
\begin{equation}
ds^2 = \frac{ \left(1 - \frac1\r\right)^{2\gamma}}{\left(1 +
\frac1\r\right)^{2\gamma}}\,dt^2 - \frac{1}{4}a_0^2 \,{ \left(1 -
\frac1\r\right)^{2-2\gamma}}{\left(1 +
\frac1\r\right)^{2\gamma+2}}\,(d\r^2 + \r^2 d\theta^2 + \r^2 \sin^2
\theta d\varphi^2)
\end{equation}
This form is manifestly flat at infinity and allows for a direct
physical interpretation of the parameter $\gamma$; indeed, for large
values of the rescaled radial variable $2\rho = a_0\r$
\begin{equation}
g_{00} = \frac{ \left(1 -
\frac{a_0}{2\rho}\right)^{2\gamma}}{\left(1 +
\frac{a_0}{2\rho}\right)^{2\gamma}} \simeq 1 - \frac{2a_0\gamma}\rho
\end{equation}
and therefore $m=a_0\gamma$ may be interpreted as the mass seen by
an observer at infinity. The standard (Schwarzschild-like) form of
the metric is also of interest.  We will obtain it in two steps:
first, introduce a radial coordinate $\b$ proportional to the scalar
field $\phi$ itself, by the relation
\begin{equation}
{\b} = \log\coth \frac r2 = \sqrt{\frac{4\pi }{1 - \gamma^2 }}
\;\;\phi(r)
\end{equation}
This change of coordinates recasts the metric in the following way:
\begin{equation}
ds^2 = e^{-2\b\gamma} dt^2 - \frac{a^2_0 e^{2\b\gamma}}{\sinh^4
\b}\left(d\b^2 + \sinh\b^2 (d\theta^2 + \sin^2 \theta
d\varphi^2)\right)
\end{equation}
A second radial coordinate defined as
\begin{equation}
R = \frac{a_0 e^{\b\gamma}}{\sinh \b}
\end{equation}
allows to rewrite finally the metric in the standard form:
\begin{equation}
ds^2 = g_{00} dt^2 - \frac{1}{f}\,dR^2 - R^2 (d\theta^2 + \sin^2
\theta d\varphi^2) \label{standardmetric}\end{equation} where the
metric components have parametric expressions in terms of the scalar
field $\tilde\phi$:
\begin{eqnarray}
g_{00} &=& e^{-2\gamma \b}, \\
 f  &=& \frac{\cosh^2(\b + \b_0)}{\cosh^2 \b_0} \equiv 1 -\frac{2m(R)}{R}\;\;\;\;\;
\makebox{with} \;\;\;\;\;\tanh \b_0= -\gamma.
\end{eqnarray}
\begin{figure}[h]
\centerline{\epsfxsize 9cm \epsfbox{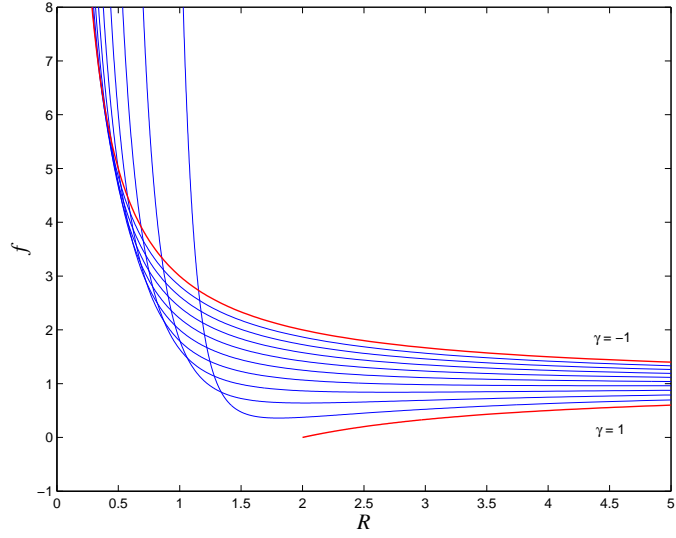}} \caption{Plot of $f(R)$
for values of $\gamma$ between -1 and 1. For $\gamma= 1$ the
Schwarzschild vacuum solution is recovered and this is the only
solution having an horizon (at $R= 2a_0$, where the corresponding
plot stops). \label{Fig.1}}
\end{figure}

\begin{figure}[h]
\centerline{\epsfxsize 9cm \epsfbox{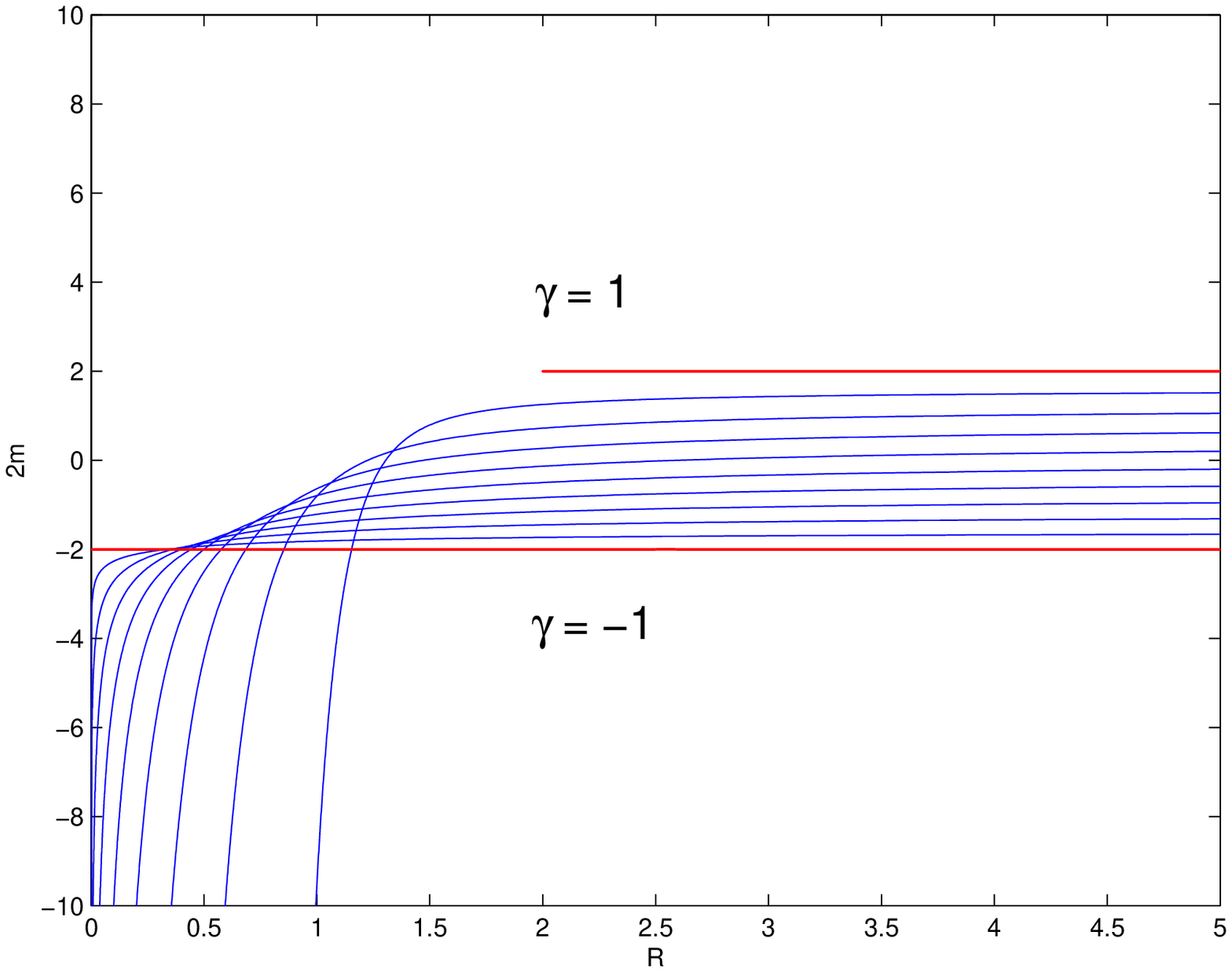}} \caption{Plot of the
mass function $2m(R)$ for values of $\gamma$ between -1 and 1. For
$\gamma= \pm1$ the mass function is constant. Outside the horizon
of the Schwarzschild solution, the mass function tends rapidly to
$m = 2\gamma a_0$.} \label{Fig.2}
\end{figure}
When $\gamma = \pm1$ the  metric (\ref{standardmetric}) reproduces
the standard Schwarzschild metric with mass $\pm 2a_0$:

\begin{equation}
ds^2 = \left(1-\frac{2a_0}{R}\right)dt^2 -
\frac{dR^2}{1-\frac{2a_0}{R}} - R^2( d\theta^2 + \sin^2\theta
d\varphi^2),\label{Schwarz}
\end{equation}
When $\gamma = 1$ the singularity at $R=0$ is hidden behind an
horizon. On the other hand, it is not difficult to check that the
when $\gamma < 1$ there is a naked singularity at $R = 0$ (see
Fig.1 and Fig. 2).

The other special case is $\gamma = 0$. In this case
\begin{equation}
ds^2 = dt^2 - \frac{dR^2}{1+\frac{a_0^2}{R^2}} - R^2( d\theta^2 +
\sin^2\theta d\varphi^2).\label{static}
\end{equation}
The corresponding manifold  is a direct product of time by the
three-dimensional static non-compact manifold with a positive
scalar curvature
\begin{equation}
R = \frac{2a_0^2}{R^4}.\label{manifold}
\end{equation}
At the point $R = 0$ the manifold (\ref{static}) has a naked
singularity.

In a completely analogous way, we now consider a static hyperbolic
metric of the following form
\begin{equation}
ds^2 = b^2(r) dt^2 - a^2(r) (dr^2 + d\chi^2 + \sinh^2\chi
d\varphi^2). \label{metrich}
\end{equation}
Einstein's equations are then written as follows
\begin{eqnarray}
&& G^t_t =  \frac{{a'(r)}^2 -{a(r)}^2  -
2\,a(r)\,a''(r)}{{a(r)}^4}= \varepsilon \label{EinK0},\\
&& G^r_r = - \frac{{a(r)}^2\,b(r) + b(r)\,{a'(r)}^2 +
2\,a(r)\,a'(r)\,b'(r)}{{a(r)}^4\,b(r)}= -\varepsilon \label{EinKr} ,\\
&& G^\chi_\chi = G^\phi_\phi =  \frac{ b(r)\,{a'(r)}^2  -
a(r)\,b(r)\,a''(r) - {a(r)}^2\,b''(r)}{{a(r)}^4\,b(r)}
\label{EinKthet}= \varepsilon
\end{eqnarray}
where again $ \varepsilon = \frac{4\pi{\phi'}^2}{a^2}. $ Now
$1/B$, defined as in  Eq. (\ref{B-def}), behaves as  an harmonic
oscillator:
\begin{equation}
\left(\frac1B\right)^{\prime\prime} + \frac1B = 0.\label{eqB}
\end{equation}
Thus
\begin{equation}
B = \frac{\gamma}{\sin r},\label{Bsol}
\end{equation}
and
\begin{equation}
A = \cot r - \frac{\gamma}{\sin r}.\label{Asol}
\end{equation}
Solutions for $a$ and $b$ easily follow:

\begin{equation}
a = \frac{a_0\sin r}{\left(\tan\frac{r}{2}\right)^{\gamma}},
\label{asol} \;\;\;\;\;\;\;\; b =
b_0\left(\tan\frac{r}{2}\right)^{\gamma},\label{bsol}
\;\;\;\;\;\;\;\; {\phi'}^2 = \frac{1-\gamma^2}{4\pi\sin^2
r};\label{scalK}
\end{equation}
again the exponent $\gamma$ should satisfy the restriction $
\gamma^2 \leq 1$. As before the values $\gamma = \pm1$ correspond to
the limiting case of empty spacetimes. When $\gamma = 0$,  $b(r)$ is
constant and the time $t$ is everywhere ticking at the same rate.

There is an interesting relationship linking the above static
spherically symmetric or hyperbolically symmetric solutions with
Kantowski-Sachs cosmologies \cite{Kant-Sachs,Xan1}. Consider, for
instance, the metric (\ref{metrich}). By interchanging the time
and the radial variables
\begin{equation}
t \leftrightarrow r \label{trans1}
\end{equation}
it becomes
\begin{equation}
ds^2 = - a^2(t)dt^2 + b^2(t)dr^2 - a^2(t)(d\chi^2 + \sinh^2\chi
d\varphi^2).\label{trans2}
\end{equation}
To remedy for the incorrect signs in front of
$dr^2$ and $dt^2$
we make the replacement
\begin{equation}
g_{\alpha\beta} \rightarrow -g_{\alpha\beta}\label{trans3}
\end{equation}
and get
\begin{equation}
ds^2 = a^2(t)dt^2 - b^2(t)dr^2 + a^2(t)(d\chi^2 + \sinh^2\chi
d\varphi^2).\label{trans4}
\end{equation}
Now the last two terms at the right-hand side of Eq. (\ref{trans4})
have wrong signs. To correct them, we make one more replacement:
\begin{equation}
\chi \rightarrow i\theta,\label{trans5}
\end{equation}
which finally produces the metric
\begin{equation}
ds^2 = a^2(t)dt^2 - b^2(t)dr^2 - a^2(t)(d\theta^2 + \sin^2\theta
d\varphi^2).\label{trans6}
\end{equation}

The indicated set of transformations also determines a map of the
components (\ref{EinK0}), (\ref{EinKr}), (\ref{EinKthet}) of the
Einstein tensor into those of a Kantowski-Sachs spherical universe
filled with the time-dependent massless scalar field $\phi(t)$,
which is realized through the substitutions
\begin{equation}
G_t^t  \rightarrow -{[G^{KS}]}_r^r \, \label{sub}
\end{equation}
\begin{equation}
G_r^r  \rightarrow -{[G^{KS}]}_t^t, \label{sub1}
\end{equation}
\begin{equation}
G_{\theta}^{\theta}  \rightarrow -{[G^{KS}]}_{\theta}^{\theta},
\label{sub2}\end{equation}

A similar set of transformations performs the transition from the
static spherically symmetric  metric to  the cosmological
hyperbolic Kantowski-Sachs metric.

The non-triviality of transformations (\ref{trans1}),
(\ref{trans3}), (\ref{trans5}) consists in the fact that they not
only exchange the radial and time variables among themselves, but
also substitute the spherical symmetry by the hyperbolical one and
vice versa.

In the hyperbolic static empty space ($\gamma=1$) the metric becomes
\begin{equation}
ds^2 = b_0^2 \tan^2\frac{r}{2}dt^2 - 4a_0^2
\cos^4\frac{r}{2}(dr^2+d\chi^2 +\sinh^2\chi d\varphi^2).
\label{metric-empty}
\end{equation}
In analogy with the spherical case, we introduce a new radial
variable
\begin{equation}
\rho \equiv 2a_0\cos^2\frac{r}{2}, \label{rho}
\end{equation}
in terms of which , the metric (\ref{metric-empty}) takes the
following ``pseudo-Schwarzschild'' form
\begin{equation}
ds^2 = \left(\frac{2a_0}{\rho}-1\right)dt^2 -
\frac{d\rho^2}{\left(\frac{2a_0}{\rho}-1\right)} -\rho^2(d\chi^2 +
\sinh^2\chi d\varphi^2),\label{metric-empty1}
\end{equation}
where again we have set $b_0 = 1$, which amounts simply to a
rescaling of the time parameter. The metric (\ref{metric-empty1})
was written down by Harrison in \cite{Harrison}, where he studied
exact three-variable empty space solutions of the Einstein
equations. There it was dubbed the degenerate solution III-9 but
its properties were not discussed, nor have they been discussed
anywhere so far to our knowledge. Here we would like to attract
the reader's attention to its peculiar properties.

To proceed further, first observe that the metric
(\ref{metric-empty1}) has the property
\begin{equation}
g_{\rho\rho} = -g^{tt},\label{Schwarz1}
\end{equation}
which it shares with the Schwarzschild metric, written in the
standard form.

Write down the geodesics equations for the radial motion valid for
both  metrics, the metric (\ref{metric-empty1}) and the
Schwarzschild one:
\begin{equation}
\frac{d^2 t}{ds^2} + 2\Gamma_{t\rho}^{t}\frac{dt}{ds}
\frac{d\rho}{ds} = 0,\label{geod}
\end{equation}
\begin{equation}
\frac{d^2 \rho}{ds^2} +
\Gamma_{tt}^{\rho}\left(\frac{dt}{ds}\right)^2 +
\Gamma_{\rho\rho}^{\rho}\left(\frac{d\rho}{ds}\right)^2 =
0.\label{geod1}
\end{equation}
Eq. (\ref{geod}) can be rewritten as
\begin{equation}
\frac{d}{ds}\left(\log\frac{dt}{ds}\right) =-
g^{tt}\frac{dg_{tt}}{d\rho} = -\frac{d\log
g_{tt}}{ds},\label{geod2}
\end{equation}
and, hence,
\begin{equation}
\frac{dt}{ds} = \frac{C_0}{g_{tt}}.\label{geod3}
\end{equation}
We fix the normalization of $t$ by choosing $C_0 = 1$.

For the purpose of solving equation (\ref{geod1}), we introduce a
new variable:
\begin{equation}
x \equiv \left(\frac{d\rho}{ds}\right)^2,\label{x}
\end{equation}
and use the relation
\begin{equation}
\frac{d^2 \rho}{ds^2} = \frac{d\rho}{ds}
\frac{d}{d\rho}\left(\frac{d\rho}{ds}\right) =
\frac{1}{2}\frac{d}{d\rho}\left(\frac{d\rho}{ds}\right)^2 =
\frac12\frac{dx}{d\rho}. \label{relation}
\end{equation}
We  need also the expressions for the Christoffel symbols:
\begin{equation}
\Gamma_{tt}^{\rho} = -\frac12 g^{\rho\rho}\frac{dg_{tt}}{d\rho} =
\frac12 g_{tt} \frac{dg_{tt}}{d\rho},\label{Chris}
\end{equation}
\begin{equation}
\Gamma_{\rho\rho}^{\rho} =
\frac12g^{\rho\rho}\frac{dg_{\rho\rho}}{d\rho} =
-\frac12\frac{d\log g_{tt}}{d\rho}. \label{Chris1}
\end{equation}
Substituting Eqs. (\ref{Chris}),(\ref{Chris1}),(\ref{geod3}) and
(\ref{x}) into Eq. (\ref{geod1}), we rewrite it in the following
form:
\begin{equation}
\frac{dx}{d\rho} - \frac{d\log g_{tt}}{d\rho} x + \frac{d\log
g_{tt}}{d\rho} = 0.\label{geod4}
\end{equation}
This equation can be immediately integrated to give
\begin{equation}
x = 1 + D_0 g_{tt}. \label{x-sol}
\end{equation}
Combining Eqs. (\ref{x-sol}) and (\ref{geod3}) one gets
\begin{equation}
\left(\frac{d\rho}{dt}\right)^2 = g_{tt}^2 + D_{0}g_{tt}^3.
\label{velocity}
\end{equation}
Formula (\ref{velocity}) is valid for both the metrics: for the
Schwarzschild and for the pseudo-Schwarzschild one
(\ref{metric-empty1}). In both cases the values $D_0 > 0, D_0 =0,
D_0 < 0$ correspond to spacelike, lightlike and timelike geodesics
respectively.

For the Schwarzschild metric the case $D_0 > -1$ corresponds to a
situation when at  spatial infinity $\rho \rightarrow \infty$ the
particle has some non vanishing velocity:
\begin{equation}
\vec{v}^2 = 1 + D_0. \label{vel}
\end{equation}
The case $D_0 < -1$ describes the situation when the particle cannot
reach spatial infinity having a turning point at
\begin{equation}
\rho_{turn} = \frac{2 a_0 D_0}{1 + D_0}. \label{turning}
\end{equation}
The value $D_0 = -1$ corresponds to the particle arriving at
infinity with  vanishing velocity.

We now turn to the study of the pseudo-Schwarzschild metric
(\ref{metric-empty1}). This metric  has a horizon at $\rho = 2a_0$,
which is in a way analogous to the Schwarzschild horizon. The
singularity in terms of the coordinate $\rho$ occurs at $\rho = 0$
and is also analogous to the Schwarzschild singularity. The main
difference lies in the fact that the metric (\ref{metric-empty1}) is
defined at $\rho < 2a_0$, i.e. inside the horizon, in contrast to
the Schwarzschild metric.

Looking at Eq. (\ref{velocity}) one can see that in the vicinity
of the horizon, when $g_{tt} \rightarrow 0$, the second term in
this equation can be omitted and the equation itself can be
integrated:
\begin{equation}
t = t_0 - \rho - 2a_0 \log(2a_0 -\rho).\label{time}
\end{equation}
Thus, the approach to the horizon, described in terms of
coordinates (\ref{metric-empty1}), takes an infinite time just as
in the Schwarzschild case. It is easy to check that as in the
Schwarzschild case the reaching of the horizon takes a finite
interval of proper time.

We now study what is going on in the vicinity of the singularity.
It is easy to see that at
\begin{equation}
\rho_{0} = \frac{2a_0}{1-\frac{1}{D_0}} \label{stop}
\end{equation}
the velocity of a massive particle vanishes and the latter cannot
reach the singularity. Curiously, there is no obstruction for
the light rays (massless particles) which fall to the singularity
$\rho = 0$.

As has already been mentioned above the metric
(\ref{metric-empty1}) is defined only at $\rho < 2a_0$. To
describe the whole manifold one can construct Kruskal-type
coordinates, analogous to those describing the Schwarzschild
manifold \cite{Kruskal}. Following a well-known algorithm (see
e.g. \cite{ABS}), one gets
\begin{equation}
ds^2 = f^2(u,v)(dv^2 - du^2) - \rho^2(d\chi^2 + \sinh^2\chi
d\varphi^2),\label{Kruskal}
\end{equation}
where
\begin{equation}
f^2(u,v) =
\frac{32a_0^3}{\rho}\exp\left(-\frac{\rho}{2a_0}\right).\label{f-def}
\end{equation}
In the  region $\rho < 2a_0$ below the horizon (see region $I$ in
Fig. 3), the relations between the Schwarzschild-type coordinates
$\rho$ and $t$  and Kruskal-type coordinates $u$ and $v$ are
\begin{equation}
u = \exp\left(\frac{\rho}{4a_0}\right)\sqrt{1-\frac{\rho}{2a_0}}
\cosh \frac{t}{4a_0},\label{Kruskal1}
\end{equation}
\begin{equation}
v = \exp\left(\frac{\rho}{4a_0}\right)\sqrt{1-\frac{\rho}{2a_0}}
\sinh \frac{t}{4a_0},\label{Kruskal2}
\end{equation}
or
\begin{equation}
u^2 - v^2 =
\exp\left(\frac{\rho}{2a_0}\right)\left(1-\frac{\rho}{2a_0}\right),
\label{Kruskal3}
\end{equation}
\begin{equation}
\frac{v}{u} = \tanh \frac{t}{4a_0}.\label{Kruskal4}
\end{equation}
Outside of horizon $\rho > 2a_0$ (see region $II$ in Fig. 3) these
relations take the following form:
\begin{equation}
u =
\exp\left(\frac{\rho}{4a_0}\right)\sqrt{\frac{\rho}{2a_0}-1}\sinh
\frac{t}{4a_0},\label{Kruskal5}
\end{equation}
\begin{equation}
v =
\exp\left(\frac{\rho}{4a_0}\right)\sqrt{\frac{\rho}{2a_0}-1}\cosh
\frac{t}{4a_0},\label{Kruskal6}
\end{equation}
or
\begin{equation}
v^2 - u^2 =
\exp\left(\frac{\rho}{2a_0}\right)\left(\frac{\rho}{4a_0}-1\right),
\label{Kruskal7}
\end{equation}
\begin{equation}
\frac{u}{v} = \tanh \frac{t}{4a_0}.\label{Kruskal8}
\end{equation}
In the regions $III$ and $IV$ the Kruskal-type coordinates are,
respectively,
\begin{equation}
u = -\exp\left(\frac{\rho}{4a_0}\right)\sqrt{1-\frac{\rho}{2a_0}}
\cosh \frac{t}{4a_0},\label{Kruskal9}
\end{equation}
\begin{equation}
v = -\exp\left(\frac{\rho}{4a_0}\right)\sqrt{1-\frac{\rho}{2a_0}}
\sinh \frac{t}{4a_0},\label{Kruskal10}
\end{equation}
and
\begin{equation}
u =
-\exp\left(\frac{\rho}{4a_0}\right)\sqrt{\frac{\rho}{2a_0}-1}\sinh
\frac{t}{4a_0},\label{Kruskal11}
\end{equation}
\begin{equation}
v =
-\exp\left(\frac{\rho}{4a_0}\right)\sqrt{\frac{\rho}{2a_0}-1}\cosh
\frac{t}{4a_0}.\label{Kruskal12}
\end{equation}
The singularity $\rho = 0$ is located at the hyperbola $u^2 - v^2
= 1$, intersecting the axis $v = 0$ at the points $u = \pm 1$.
Remember, that the Schwarzschild singularity is located at the
hyperbola $v^2 - u^2 = 1$. Thus, the Kruskal diagram for the
pseudo-Schwarzschild manifold is rotated by $\pi/2$ with respect
to the familiar Kruskal diagram for the Schwarzschild manifold. In
other words, in contrast to the Schwarzschild singularity which is
spacelike, the pseudo-Schwarzschild singularity is timelike.
When speaking of the Kruskal diagram for the pseudo-Schwarzschild
manifold, it is necessary to keep in mind that its two-dimensional part,
which is not drawn explicitly, is hyperbolic.

Looking at formula (\ref{metric-empty1}) we note an important
analogy of the pseudo-Schwarzschild metric to the Schwarzschild
one (\ref{Schwarz}). Whereas the Schwarzschild geometry is static
outside the horizon and time-dependent inside, it happens the
other way around with the pseudo-Schwarzschild one. Indeed, the
coordinates $\rho$ and $t$ are respectively spacelike and timelike
in the region $0 < \rho < 2a_0$, whereas in the region $\rho >
2a_0$ the coordinate $\rho$ is timelike and the coordinate $t$ is
spacelike.

There is also another difference between the Schwarzschild and
pseudo-Schwarzschild manifold. In the Schwarzschild case, as is
well known one has two types of timelike radial geodesics,
depending on the value of the parameter $D_0$: those whose motion
is finite ($D_0 < -1$) and those whose motion is infinite ($D_0
\geq -1$), In terms of the Kruskal diagram of Fig. 3, the former
travel from region $IV$ through region $I$ (or $III$) to region
$II$, i.e. they begin their motion from the (white hole)
singularity $S_{IV}$, cross the horizon, reach a turning point
where the radial coordinate acquires a maximum value, then cross
the horizon again and fall onto the (black hole) singularity
$S_{II}$. Instead, the geodesics corresponding to $D_0 > -1$
either arrive from spatial infinity of region $I$, cross the
horizon and fall onto the singularity $S_{II}$ (black hole), or
leave the singularity $S_{IV}$, cross the horizon and travel to
spatial infinity in region $III$ (white hole), (see Fig. 3).

\begin{figure}[h]
\epsfxsize 6cm \epsfbox{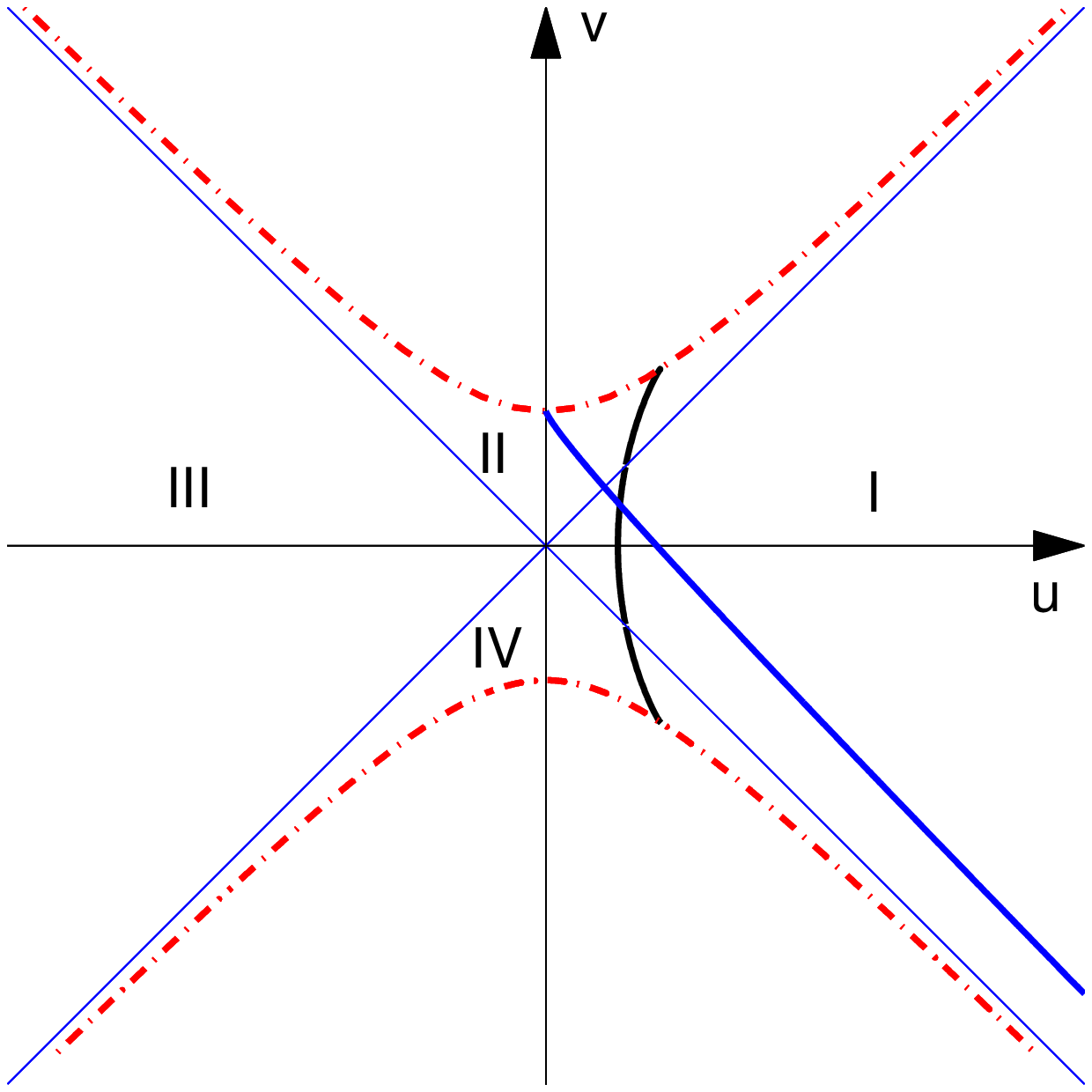} \hfill \epsfxsize 6cm
\epsfbox{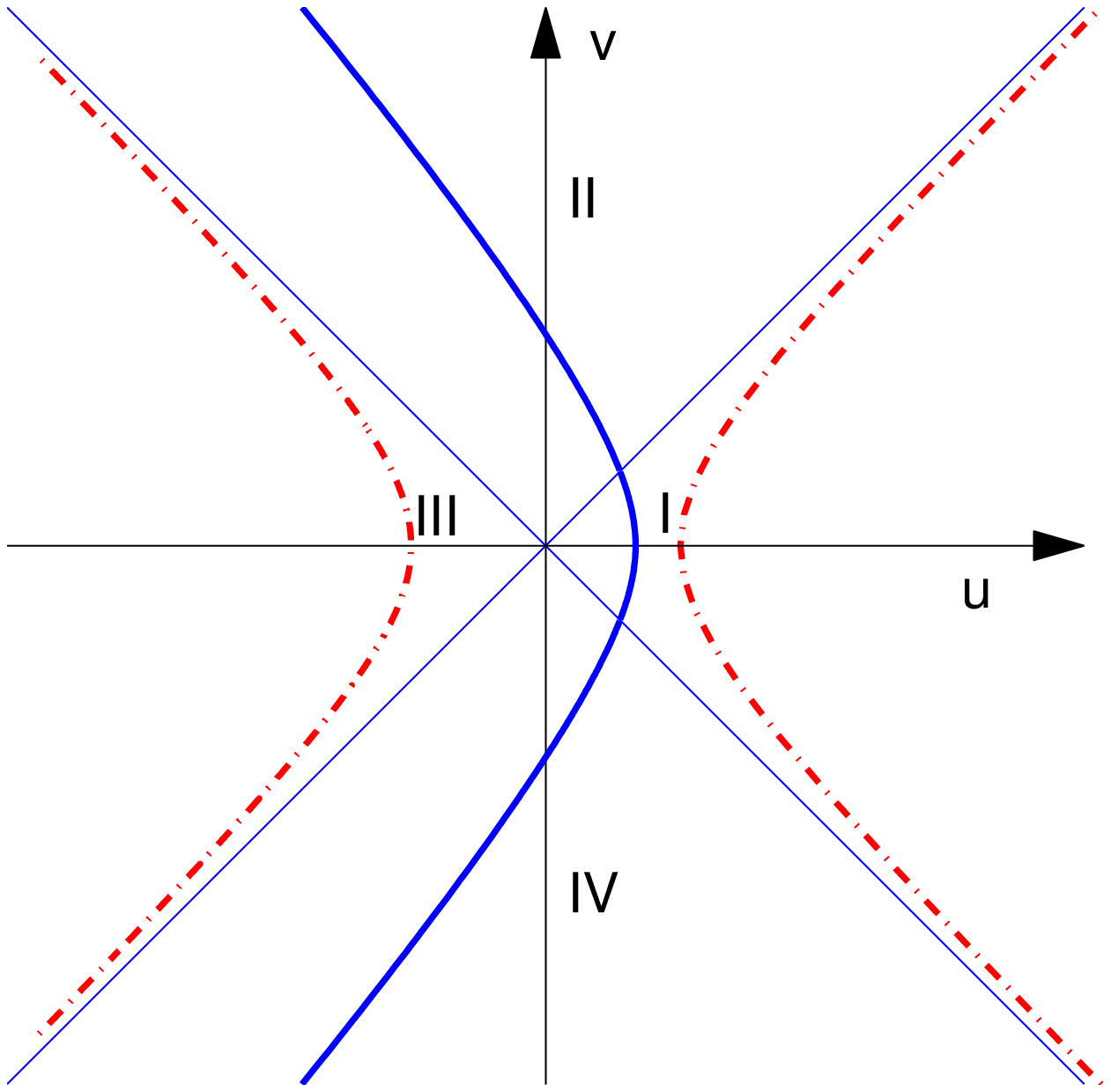}  \caption{The Kruskal diagrams for the
Schwarzschild (on the left) and pseudo-Schwarzschild (on the right)
manifolds.} \label{Fig.3}
\end{figure}

By contrast, in the pseudo-Schwarzschild manifold there exists
only one type of timelike geodesics: a test particle travels along
the geodesics from spatial infinity of region $IV$, cross the
horizon entering region $I$ (or $III$), reaches the turning point
(\ref{stop}), (where the distance of the particle from the
singularity attains its minimum value), cross again the horizon
entering region $II$ within which it travels to spatial infinity
(see Fig. 3).

In analogy to the Schwarzschild case \cite{Chandra} the equation
of motion for the radial timelike geodesic (\ref{velocity}) in the
pseudo-Schwarzschild geometry can be integrated parametrically. To
achieve this, using the relation (\ref{stop}) we rewrite equation
(\ref{velocity}) in the form
\begin{equation}
\frac{d\rho}{dt} =\left(\frac{2a_0}{\rho}-1\right)
\sqrt{1-\frac{(2a_0-\rho)\rho_0}{(2a_0-\rho_0)\rho}}
\label{geod-pseud}
\end{equation}
and  introduce the parameter $\eta$ as
\begin{equation}
\rho = \rho_0 \cosh^2\eta. \label{eta}
\end{equation}
Then Eq. (\ref{geod-pseud}) can be integrated to give, up to an
additive constant,
\begin{eqnarray}
&&t = -\rho_0\sqrt{1-\frac{\rho_0}{2a_0}}\eta -
4a_0\sqrt{1-\frac{\rho_0}{2a_0}}\eta -\frac{\rho_0}{2}
\sqrt{1-\frac{\rho_0}{2a_0}}\sinh 2\eta \nonumber \\
&&+2a_0\log\left|\frac{(e^{2\eta}-e^{-2\eta_H})
(e^{2\eta_H}-1)}
{(e^{2\eta_H}-e^{2\eta})(1-e^{-2\eta_H})}\right|,\label{time-sol}
\end{eqnarray}
where
\begin{equation}
\eta_H = arccosh \sqrt{\frac{2a_0}{\rho_0}} \label{eta-cros}
\end{equation}
is the value of $\eta$ which corresponds to
the crossing of the horizon.

It is easy to see that when $\eta \rightarrow \eta_H$, the
parameter $t$ tends to infinity. Then, with the growth of the
parameter $\eta$ the parameter $t$ is decreasing and at $\eta
\rightarrow \infty, t \rightarrow -\infty$. Thus, just like in the
case of the Schwarzschild metric the crossing of the horizon
requires an infinite coordinate time.

We can also find the dependence of the proper time $\tau$ on the
parameter $\eta$. Using Eq. (\ref{geod-pseud}) and the relation
$d\tau^2 = g_{tt}dt^2 +g_{\rho\rho}d\rho^2$ we get
\begin{equation}
\frac{d\rho}{d\tau} = \frac{2a_0}{\rho_0} - \frac{2a_0}{\rho}.
\label{tau}
\end{equation}
Using the parametrization (\ref{eta}) one finds
\begin{equation}
\tau  = \frac{\rho_0}{2}\sqrt{\frac{\rho_0}{2a_0}}(\sinh 2\eta +
2\eta). \label{tau1}
\end{equation}
The proper time which a massive particle needs to cover the
distance from the point $\rho_0$ to the horizon is finite and
given by the formula
\begin{eqnarray}
&&\tau_H = \frac{\rho_0}{2}\sqrt{\frac{\rho_0}{2a_0}}(\sinh
2\eta_H
+ 2\eta_H)\nonumber \\
&&= \rho_0\sqrt{\frac{2a_0}{\rho_0}-1} +
\rho_0\sqrt{\frac{\rho_0}{2a_0}}arccosh\sqrt{\frac{2a_0}{\rho_0}}.
\label{tau2}
\end{eqnarray}

We note that
$$
\tau_H(0) = \tau_H(2a_0) = 0,
$$
as expected. Therefore there exists a value of $\rho_0$ at which
$\tau_H$ is maximum. The equation $\frac{d\tau_H(\rho_0)}{d\rho_0}
= 0$ is transcendental and cannot be solved analytically. The
graph of the function $\tau_H(\rho_0)$ is depicted in Fig. 4. One
sees that the maximum value of $\tau_H$ is achieved at $\rho_0
\approx 0.61\times 2a_0$.
\begin{figure}[h]
\epsfxsize 6cm \epsfbox{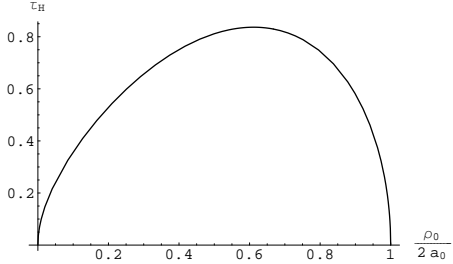} \caption{$\tau_H$
dependence on the ratio $\frac{\rho_0}{2a_0}$.} \label{Fig.4}
\end{figure}

Now consider the motion outside of the horizon. In this region the
time and radial coordinates change their roles. Introducing the
notations $\tilde{t}$ and $\tilde{\rho}$ for new timelike and
spacelike coordinates one writes the metric outside the horizon in
the following form:
\begin{equation}
ds^2 = \frac{d\tilde{t}^2}{1-\frac{2a_0}{\tilde{t}}}
-d\tilde{\rho}^2\left(1-\frac{2a_0}{\tilde{t}}\right)-
\tilde{t}^2(d\chi^2 + \sinh^2\chi d\varphi^2). \label{outside}
\end{equation}
This metric is non-stationary  and when $\tilde{t} \rightarrow
\infty$ it becomes the Minkowski metric. To be precise it
asymptotically tends to the Minkowski metric, written in a rather
particular way: it represents a direct product of the  line
$\tilde{r}$ times the (2+1) - dimensional Milne manifold.

The geodesic equation for massive particles can be reduced to the
following form:
\begin{equation}
\left(\frac{d\tilde{\rho}}{d\tilde{t}}\right)^2 = \frac{v^2
g_{tt}^3}{1 - v^2 + v^2 g_{tt}}, \label{outside1}
\end{equation}
where $v$ is the asymptotic value of the velocity
$d\tilde{\rho}/d\tilde{t}$ when $\tilde{t} \rightarrow \infty$.

One can find the relation between the velocity parameter $v$ and
the parameter $\rho_0$ of the minimal distance between a particle
under consideration and the singularity, using Eqs. (\ref{eta})
and (\ref{time-sol}). Indeed,
\begin{equation}
v = \lim_{\tilde{t} \rightarrow \infty}
\frac{d\tilde{\rho}}{d\tilde{t}} = \lim_{\eta \rightarrow \infty}
\frac{dt}{d\rho} = \sqrt{1-\frac{\rho_0}{2a}}.
\label{r-v-relation}
\end{equation}

We now come back to the Kantowski-Sachs cosmological solutions for
both the spherically symmetric  and  the hyperbolic  universes and
consider their empty space limits $\gamma = 1$. The corresponding
metrics are respectively
\begin{equation}
ds^2 = 4a_0^2\cos^4\frac{t}{2} dt^2 - \tan^2\frac{t}{2}dr^2 -
4a_0^2\cos^4\frac{t}{2}(d\theta^2 + \sin^2\theta d\varphi^2),
\label{KS-empty-sp}
\end{equation}
\begin{equation}
ds^2 = 4a_0^2\cosh^4\frac{t}{2} dt^2 - \tanh^2\frac{t}{2}dr^2 -
4a_0^2\cosh^4\frac{t}{2}(d\chi^2 + \sinh^2\chi d\varphi^2).
\label{KS-empty-hyp}
\end{equation}
The metrics of the empty Kantowski-Sachs universes, written in the
form (\ref{KS-empty-sp}) and (\ref{KS-empty-hyp}) do not describe
complete manifolds just like their static analogs. Making the
transformation
\begin{equation}
\tilde{t} = 2a_0\cos^2\frac{t}{2} \label{KS-empty-sp1}
\end{equation}
one sees that the metric (\ref{KS-empty-sp}) describes the
internal part of the Schwarzschild world (as was noticed in paper
\cite{Kant-Sachs}). Completing the manifold we construct the
external part of the Schwarzschild world, which is static.

Similarly, making the transformation
\begin{equation}
\tilde{t} = 2a_0 \cosh^2\frac{t}{2} \label{KS_empty-hyp1}
\end{equation}
we get the metric, describing the external part of the
pseudo-Schwarzschild world which is nothing but (\ref{outside}),
while its completion gives the static pseudo-Schwarzschild
geometry (\ref{metric-empty1}) below the horizon.

\begin{picture}(700,500)
\put(0,0){\framebox(125,60){\shortstack{external part of the
pseudo-
\\Schwarzschild manifold}}}
\put(300,0){\framebox(125,60){\shortstack{internal part of the
pseudo-
\\Schwarzschild manifold}}}
\put(295,30){\vector(-1,0){165}} \put(130,30){\vector(1,0){165}}
\put(145,35){completion of the manifold}
\put(0,150){\framebox(125,60){\shortstack{Kantowski-Sachs hyperbolic\\
universe}}} \put(150,180){``duality'' transformation}
\put(140,185){\vector(1,1){155}} \put(140,185){\vector(-1,-1){10}}
\put(285,185){\vector(-1,1){155}} \put(285,185){\vector(1,-1){10}}
\put(60,365){\vector(0,1){80}} \put(360,365){\vector(0,1){80}}
\put(300,150){\framebox(125,60){\shortstack{static
hyperbolic\\universe}}} \put(60,145){\vector(0,-1){80}}
\put(360,145){\vector(0,-1){80}}
\put(0,300){\framebox(125,60){\shortstack{Kantowski-Sachs spherically\\
symmetric universe}}}
\put(300,300){\framebox(125,60){\shortstack{static spherically\\
symmetric universe}}}
\put(0,450){\framebox(125,60){\shortstack{internal part of \\the
Schwarzschild manifold}}}
\put(300,450){\framebox(125,60){\shortstack{external part of \\the
Schwarzschild manifold}}} \put(130,480){\vector(1,0){165}}
\put(295,480){\vector(-1,0){165}} \put(145,485){completion of the
manifold} \put(370,100){\shortstack{empty\\space\\limit}}
\put(70,100){\shortstack{empty\\space\\limit}}
\put(370,390){\shortstack{empty\\space\\limit}}
\put(70,390){\shortstack{empty\\space\\limit}}
\end{picture}

The interrelations amongst different cosmological and static
solutions of the Einstein equations in the presence of massless
scalar field and their empty space limits are represented in the
scheme above.

We conclude by noting that the pseudo-Schwarzschild metric can be
obtained by applying an analogue of Birkhoff theorem to a vacuum
hyperbolically symmetric solution of the Einstein equations. Indeed,
by mimicking the proof of Birkhoff theorem one can show that the
most general vacuum hyperbolically symmetric metric is static and is
equivalent to the internal pseudo-Schwarzschild solution
(\ref{metric-empty1}).

A.K. was  partially supported by  RFBR, grant No. 05-02-17450. U. M.
and A. K. thank the Service de Physique Theorique of CEA-Saclay for
hospitality and support.


\begin{thebibliography}{99}
\bibitem{Schwarz}
Schwarzschild K 1916 {\it Sitzungsber.Preuss.Akad.Wiss.Berlin
(Math.Phys.)} 189; 424
\bibitem{Tolman}
Tolman R 1939 {\it Phys. Rev.} {\bf 55} 364
\bibitem{Op}
Oppenheimer J R and Volkoff G M 1939 {\it Phys. Rev.} {\bf 55} 374
\bibitem{Xan}
Xanthopoulos B C and Zannias T 1989 {\it Phys. Rev.} D {\bf 40}
 2564
\bibitem{Kant-Sachs}
Kantowski R and Sachs R K 1966 {\it J. Math. Phys.} {\bf 7}  443
\bibitem{Xan1}
Xanthopoulos B C and Zannias T 1992 {\it J. Math. Phys.} {\bf 33}
 1415
\bibitem{Harrison}
Harrison B K 1959 {\it Phys. Rev.} {\bf 116}  1285
\bibitem{Kruskal}
Kruskal M D 1960 {\it Phys. Rev.} {\bf 119}  1743
\bibitem{ABS}
Adler R, Bazin M and Schiffer M 1975 {\it Introduction to General
Relativity} (New York: McGraw-Hill)
\bibitem{Chandra}
Chandrasekhar S 1983 {\it The Mathematical Theory of Black
Holes}\\
(New York : Oxford University Press)
\end{thebibliography}
\end{document}